Article

# Human Cognition Surpasses the Nonlocality Tsirelson Bound: Is Mind Outside of Spacetime?

Stuart Kauffman[1] and Sudip Patra[2]

[1]Emeritus Professor of Biochemistry and Biophysics, University of Pennsylvania, stukauffman@gmail.com

[2]Associate Professor OP Jindal Global University, Founding member CEASP India; Research Fellow, The Laszlo Institute of new paradigm research, Italy. spatra@jgu.edu.in

## Abstract

Recent experimental studies on human cognition, particularly where non-separable or entangled cognitive states have been found, show that in many such cases Bell or CHSH in-equalities have been violated. The implications are that greater non-local correlations than allowed in quantum mechanics (often known as the Tsirelson bound), are found in human cognition. However, it is also evident that surpassing of Tsirelson limit within relativistic physical Spacetime seems impossible. Tsirelson limit is not guaranteed by no-signaling condition, but some deeper feature like in-distinguishability of identical quantum particles might be related to such an upper bound in quantum physics, within relativistic Spacetime. We propose in the current paper that a non-local theory of mind is needed in order to account for the empirical findings. This requires a foundationally different approach than the extant 'quantum-like' approach to human mind. To account for the surpassing of the Tsirelson bound we propose abandoning the constraint of no-signaling that depends upon spacetime. Thus, we ask; 'Is mind outside spacetime?' We discuss a candidate theory of quantum gravity based on nonlocality as fundamental that may accord with our proposal. We are led to suggest a new 6 - part ontological framework linking Mind, Matter, and Cosmos.

Key words: nonlocality, no-signaling, Bell inequalities, Tsirelson/Cirelson bound, PR boxes, Cognition, quantum gravity, six-part framework

## Introduction: Is 'mind' outside physical Spacetime?

Nonlocality has baffled us since the birth of modern science. For example, in Newton's gravity framework, we have action at a distance, and Newton himself did not want to forward any 'explanation' of the same by stating, "Hypothesis non-fingo". Later with the advent of Special Theory of relativity (SR), and then the General Theory of Relativity (GR), Einstein nearly singlehandedly challenged the age old concepts of space and time, proposing the bold and beautiful concept of spacetime, where continuity of action (COA)plays a central role. COA holds that if a spacetime event A has to influence another spacetime event B then it also has to influence any closed 3 surface between them. Hence no-signaling, or that there is an ultimate limit of signaling between spacetime events, which happens to be the speed of light in vacuum, became the foundational physical constraint for any sound theory of Physics. The orthodox quantum mechanics (QM) which emerged from intense discussions in Solvay conferences [1] and later known as Copenhagen version, was, however, still based on a Newtonian space and time background. Later, with the emergence of quantum field theory (QFT) there has been an uncomfortable coexistence of SR and QM. The holy grail of modern Physics has been to construct unified field theories, and particularly quantum gravity (QG), as the cherished unification of GR with QM. However, in all of these numerous attempts, nonlocality has been a recurrent problem. Even different interpretations of QM, starting from the collapse of the wave function to different alternative theories like Bohmian mechanics [2] or spontaneous collapse of wave function [3] have been riddled with different forms of non-localities. Very recently, different approaches to QG [4] would presume to hold nonlocality as fundamental, which is radical. Spontaneous collapse models or dynamic collapse models have attempted to resolve the measurement problem by introducing a collapse operator in the Schrödinger equation, for example in GRW [5] where a probability of such a stochastic collapse is small in case of single particles but grows exponentially in case of many body systems. Hence, the attempt has been to resolve the incompleteness or inconsistency problems in orthodox QM, for example, how in the same framework both deterministic and unitary Schrödinger evolution and random collapse of wave function due to 'measurement' can be accommodated. However, very recent work, (for example see 'consciousness and quantum mechanics' [6] and [7] now says that any "physically causal" theory for measurement is almost ruled out.

There are also physically "acausal" accounts of measurement. Here we refer to the recent consciousness induced collapse framework of Chalmers and McQueen [8] where phenomenal consciousness plays the role of a superposition resistant, hence definite consciousness state that result in "collapse". More recently Kauffman and Roli [9] and Kauffman and Radin [10] have utilized Heisenberg's interpretation of quantum mechanics in terms of ontologically real Potentia, Res potentia, and ontologically real Actuals, Res extensa, where actualization converts Possible to Actuals. This interpretation does not inherit the mind-body problem because Potentia are not substances. In turn this approach suggests a natural role for "mind" in actualizing quantum potentia, hence "collapsing the wave function. At this point, data supporting this hypothesis with respect to work using the two-slit experiment are strongly supportive at 6.49 Sigma, or $4 \times 10^{-11}$ [10]. We shall base our own discussion on Heisenberg's interpretation.

In addition to Heisenberg's "potentia" interpretation, workers have studied several other non-realist frameworks, where the wave function is not ontological, but rather a tool for computing probabilities for epistemological updates of knowledge state of observers. Two alternative strongly emerging interpretations of QM are relational QM and QBism. Relational QM holds that QM, or reality for that matter, is not described by quantum states, but rather by relations among observables. This is a fact ontology (for more details about "relative" and "stable" facts, we can refer to seminal literature. [11]). QBism agrees on placing a central role on phenomenology or subjective experiences, where QM is he navigation tool for any user (rather than defining who is the user) to make optimal decisions. In QBism relations between the elements of the framework are objective, such that every agent would agree. We differ from these frameworks in that these frameworks are largely based on the locality of physical spacetime, but then they face nonlocality problems.

One approach to surpassing the Tsirelson bound is found in PR box worlds (Popescu [12] a modern review of Tsirelson bound can be found in Stuckey et al. [13]) that allow for greater than QM non-local correlation limit the Tsirelson bound. However, the PR box worlds correspond to no physical model of a universe, [12]. Hence, a related question raised earlier was whether QM is the only theory where there is a co-existence of nonlocality in the sense of Bell inequalities violation and relativistic causality.

We propose in this paper that if we need to include mind or cognitive aspects in the foundational frameworks of nature, then we need to have nonlocality as the central feature. In this paper then, we explore our framework of non-local mind or cognition and are led to our proposal that "mind" is not in spacetime. By proposing that mind is not in spacetime, we can naturally eliminate the requirement for Continuity of Action, hence no-signaling, that makes sense only within a framework of spacetime. By proposing that mind is not in spacetime, mental events that are in spacetime but surpass the Tsirelson bound can be explained.

In order to begin to make sense of the concept that "mind is not in spacetime, but conscious events are in spacetime", we are led to propose a novel 6 part ontological framework linking Mind, Matter, and the Cosmos. The grounds for this novel framework are tentative, but we hope worthy of consideration and are testable in part.

### Section 1. Cognition beyond the Tsirelson Bound

1.2 Cognitive science experiments
Aerts et. al [14] have pioneered the study of non-separable states in individual minds or cognition. This includes how different concepts are combined. Such concept combinations in individual minds can be re-formulated as non-separable states. In technical language, these are intra state entanglements, which would mean coupling of different degrees of freedoms of a single system. Here these are individual minds, and the data can be expressed through inequalities such as CHSH as we explain in section 2. The statistical values or values obtained from ensembles of 'minds' of participants in such cognitive experiments can be inputted as inequalities. The results have demonstrated a clean violation of the 'non-local' correlation bound which occurs in QM.

Such a tight bound is, as noted, called a Tsirelson bound that is maximal and characteristic for QM. The same authors also provide the statistical significance of their results. Aerts, Arguelles and colleagues [15,16,17] have claimed a statistical significance of p values ranging 0.001-0.005, which is strong enough to suggest, but not yet prove, the viability of their results.

Aerts et. al (op. cit.) also adopts a Hilbert space framework, but their strategy is of 'reverse engineering', i.e. to start with the empirical results, and then describe such results by a suitable state space modelling, where the state space can be either Hilbert space or a larger Fock space. Thus, the usage of CHSH inequalities is statistical in nature, since such inequalities are general. Maximal algebraic violation of such inequalities can be beyond the Tsirelson bound, but when Hilbert space is the state space then a tight upper bound comes up as a constraint.

Given the above points, Aerts et al.'s claim of greater than Tsirelson bound violation in cognitive experiments raises several questions. For example, can any or all Hilbert space formulations account for such super quantum correlations? Aerts et al. have responded by suggesting that an entanglement that they consider is of a more complex nature, i.e. entanglement both in states as well as measure, might account for super violations. We assess this approach below.

We stress again that any Hilbert space formulation of quantum mechanics implies a tight Tsirelson bound. And we stress again that the Hilbert space formulation is stated in a background spacetime with "no-signaling" and continuity of action, hence "locality".

## Section 2. Nonlocality: Implications for QM

2.1 Nonlocality in QM

Here we remind ourselves of the seminal contribution of John Bell [18, 19] and state the basic requirements for Bell factorization conditions, upon which the celebrated Bell inequalities or later CHSH inequalities are based. Based on continuity of action, the following three assumptions are required for establishing Bell factorization.

Statistical Independence: a. $\rho(M) = \rho(\mu)$, where $\mu$ denotes the local hidden variable, and M stands for measurement settings of apparatuses for different space-like separated agents.

b. Output independence: $\rho_{ab}(a, b, \mu) = \rho_a(x_a |a, b, \mu)\rho_b(x_b |a, b, \mu)$, subscripts a and b stands for different agents, namely, Alice and Bob, x's are outcomes at their ends and a, and b are inputs at their ends respectively.
c. Parameter independence: $\rho_a(a, b, \mu) = \rho_a(a, \mu)$, $similarly$ $\rho_b(a, b, \mu) = \rho_b(b, \mu)$ Hence, in conjunction of the three assumptions we have the Bell factorization, $\rho_{ab}(a,b,\mu) = \rho_a(a, \mu). \rho_b(b, \mu).$ (1).

Bell factorization is a general condition based on the local realism assumptions (COA to be precise), which is violated by different theories in different ways. For example, QM violates Bell factorization by violating output independence but keeping statistical independence and parameter independence. Super Deterministic theory violates the same by violating Statistical independence, while keeping the other assumptions. And Cavalcanti and Wiseman [20] have showed how Bell factorization can be derived from conjunction of local 'signalism' and predictability.

In the form of CHSH, we have two space-like separated agents, Alice and Bob, where say the measurement settings in Alice's end are {a, a'} and Bob's end are {b, b'}, and all results are dichotomous (say, +/- 1). Here we define the correlation function as : $C(a,b) = P_{ab}(+1,+1) + P_{ab}(-1,-1) - P_{ab}(1,-1) - P_{ab}(-1,+1)$ (2)

Hence, we have the CHSH inequality as $CHSH$ = C(a, b) + c(a, b' ) + c(a' , b) − c(a' , b' ) (3).

Hence CHSH has different upper bounds for different underlying theories. For example, for a local deterministic theory (COA is the requisite here) we would always have as

|$CHSH$|≤2.  For QM the maximum violation of the above limit would take place when

C(a, b) =  (a, b'') =  c(a'', b) =− c(a', b') =   √2/2 , hence  this gives the Tsirelson (T bound from now) bound of |$CHSH$|≤2√2.  However algebraically it is possible that we have:

c(a, b) =  c(a', b') =  c(a', b') =− c(a', b') =  1, hence making the maximal upper bound as 4.

2.2 Different forms of non-separable states: QM and beyond

We mention here that generally composite systems in QM can be represented as tensor products of states belonging to different Hilbert spaces, such that the total Hilbert space of the composite system is a tensor product of such Hilbert spaces. This context is called product states. In addition, we recall that a Tensor product space is strictly larger that space of direct sums, hence this context also captures 'quantum-holism'. Now the typical definition of an intersystem entanglement is when the composite system state cannot be defined as simple tensor products of subsystem states. Intersystem entanglement is most discussed in QM literature, since that is what generates non-local correlations. In an entangled state the whole is always in a pure state, whereas parts are not in pure states, this is the classical Schrödinger way of denoting entanglement. Again, as we have stated earlier, maximally entangled states (often called as Bell states) can violate CHSH maximally until the T bound.

However, intra-system entanglement, defined as coupling between multiple degrees of freedom of the same system, is also discussed widely. Particularly in the classical electromagnetism literature authors, Ghose and Mukherjee [21], have observed widely that intra system entanglement, for example coupling between path and polarization states of a vortex beam, can produce such non-separable states (at times called Shimony-Wolf states) which can generate violations of CHSH inequalities. Authors, for example, Khrennikov [22, 23] has suggested that intra and inter system entanglements is the main difference between quantum and so-called 'classical' entanglement.

Multipartite nonlocality: Traditionally Bell tests or CHSH tests are bi-partite nonlocality tests, there have been several modifications though, for example GHZ states or W states, which extends frameworks for many body entanglement. In our previous framework [24] we start with a multipartite entanglement state. However, its only recently, Bancal et al. [25], that a suitable mathematical framework is being built. Here we refer to the basic tenets of such a framework, since this might be harnessed in the framework we suggest here.

We mention here that nonlocality is a recurrent feature for many-body systems too (see for example in Bancal et al. [25], for example if we consider a tripartite system, with say each subsystem possessed by Alice, Bob and Charlie who are spatially separated. Say Alice, Bob and Charlie's experimental set ups are X, Y and Z respectively and outcomes of experiments are a, b and c respectively (binary outcomes for simplicity).
$P(XYZ) = \sum_l q_l P(X)P(Y)P(Z)$ where q's are bounded by 0 and 1 sum to unity, then the sum represents local correlations, where the subscript l is for underlying hidden variables.

At times, such contexts are also called hybrid nonlocality. In another related literature (Bennet et al [26] as one seminal work in this direction) nonlocality without entanglement is theoretically proposed, and later experimentally verified. We refer to these studies to seek further support for our assertion that nonlocality is a more universal and genuine feature of reality. We also are aware of studies differentiating between genuine nonlocality and direct influences (see for example Atmanspacher et al. [27].

2.3 Attempts to fit the evidence for nonlocality within the framework of a backgroundSpacetime.

In the last century intense debate on nonlocality, or more precisely what nonlocality should mean given relativistic spacetime, was a major debate, and is still continuing. Thenonlocality debate has also thrown deeper light on the foundational thinking on QM.We observe here that the axioms of special theory of relativity (COA) or consequently fundamental limit for speed of signaling between spacetime events, and the equivalenceof inertial reference frames) seems to be elegant and physically based. However, the axioms of QM seem to be mathematical with no clear physical basis.

Aharonov and Bohm [28], and later Popescu [12], and independently, Shimony [29] have proposed that QM has to be compatible with relativistic causality, hence with Continuity of Action, COA. The efforts of the authors mentioned showed thatnon-local correlations, for example in an EPR set up, can be compatible with relativistic causality if and only uncertainty of outcomes of measurements is fundamental. Or in other words the effect of a cause here is uncertain. (Thus, counterfactuals are required). Aharonov was the first to propose 'modular' quantum variables, that are non-local in spacetime due to non-local relativistic phases, and they have optimal uncertainty for no-signaling. Shimony amusingly observed the whole affair is 'passion at a distance'.

To sum up, it is well known now that only no-signaling condition is not sufficient to guarantee Tsirelson bound in quantum physics (Amorim [30], hence it is suggested that nonlocality is a stronger condition than entanglement. Nonlocality implies entanglement and, or non-compatibility of local measures, but not the other way round. Suggestion here is that more deeper features like in-distinguishability of identical quantum particles or Qubit systems is one source of Tsirelson bound.

2.4 Attempts to surpass the Tsirelson bound in formal models.

Based on the dense PR box literature, there have been many attempts to make super quantum correlations (violating T bound) compatible with relativistic causality, or COA in general, for example, Popescu, [12]. Related questions have been whether QM is theonly possible theory where non-local correlation and no-signaling co-exists? [12]. Or why QM does not exhibit greater nonlocality? (Linden et al. [31] for example). We further observe that there have been efforts in the line of including communication complexity, and or, information causality to eradicate super quantum correlations. We also note that super correlations or greater than T bound violations are possible in configuration spaces with very particular properties. Overall, there has been an attempt to make violations of Bell inequalities (not super correlations) compatible with relativistic causality, but it is far from clear what would be the implications for super correlations for a locality criterion.

As we explore below, how violations of Bell / CHSH or even super correlation results have been observed in cognitive experiments. We also note that some authors (Khrennikov [22.23] have observed that if the observables in a particular theory cannot be represented by Hermitian operators, there might not be any T bound constraint.

### Section 3. Is Spacetime Fundamental?

3.1 Zeilinger and Information: It is important to stress that several authors are exploring the idea that spacetime is not fundamental. In particular, Zeilinger has proposed that "information" is fundamental and somehow spacetime emerges from "information" (see for example Zeilinger's seminal works since 1998 [32]. We note a central issue, "information" itself implies "possibilities" that are not either true or false. Consider Shannon information and the source. A given bit string, say (11111) can carry no information unless one of the bits can, counterfactually, be 0. That is, it must be possible that one of the bits is 0 not 1. Thus, the very concept of "information" requires more than one simultaneously possible state of the universe.

3.2 Res potentia and Res extensa linked by measurement: In the current article, we base our approach on Heisenberg's interpretation of the quantum state as "potentia standing ghost – like between an idea and reality". One of us [33] has developed Heisenberg's interpretation as "Res potentia" ontologically real Possibles, and Res extensa, ontologically real Actuals. Possibles do not obey Aristotle's law of the excluded middle and law of noncontradiction, so are neither 'true' nor 'false'. This allows "Potentia" to explain quantum superpositions: "Schrödinger's cat simultaneously is possibly alive and possibly dead." This is not a contradiction.

Potentia are non-spatial in nature but ontologically real. By contrast Actuals do obey Aristotle's two laws, so are either true or false. All of Classical physics is based on such true false Boolean variables. Given the concept of Res potentia, one of us, Kauffman [4] has explored a new approach to quantum gravity that takes nonlocality to be fundamental. Non locality taken as fundamental implies that spacetime is not itself fundamental but must somehow arise from the behaviors of entangled coherent, hence non local, quantum variables. Then non-local entangled coherent quantum variables, "Res potentia" are not in spacetime. They are Potentia not in spacetime.

3.3 Mind and the Quantum Vacuum: One natural interpretation of the line of thought above is that the quantum vacuum consists precisely in non-local entangled quantum coherent variables. Given the above, a natural proposal is that 'mind' – non-spatial, is identical or related to the quantum vacuum. We here both propose this identity and explore its potential validity.

A first implication of the proposed identity of mind and the quantum vacuum is that both are outside of spacetime. This is a possible step to explaining Aert's results. To do so, we need to show that surpassing the Tsirelson bound is straight forward if mind is outside of spacetime. In this case we can abandon no-signaling and continuity of action. We show this next. But there is a further issue, Aerts et. al data concern experiences of humans and those experiences are in spacetime. Powerful recent arguments now strongly suggest that conscious experiences (phenomenal nature) arise upon collapse of the wave function, hence, qualia are in spacetime. And further remarkable evidence now clearly shows that we can purposefully actualize the wave function. A responsible free will is not ruled out. We address all this below. These recent results and claims will be part of our proposed 6 part framework introduced below.

### Section 4. Surpassing the Tsirelson Bound if Mind is Outside of Spacetime

Aerts et al. [15 -17] themselves have attempted to justify the super quantum correlation values obtained in their 'concept-combination' experiments based on complex entanglement nature in their experimental settings, given that the configuration space of mind is a high dimensional Hilbert space. However, the standard belief, going back to Popescu [12], has been that the maximum limit of 'nonlocality' allowed in a Hilbert space is the bound.

Our perspective is not to justify the super violations based on the complexity of entanglement (both in states and in measurements), since there have been critiques of this line of argument by suggesting that if the 'marginal selectivity' rule is also violated along with Bell inequalities, which Aerts et al. observes, then there can be contaminations in testing for Bell violations. Hence we propose the six part framework, where our definition of mind need not be constrained by any physical locality condition.

### Section 5. Mind and Qualia – Collapse of the Wave Function

Recently Chalmers and McQueen [8], who have been very skeptical about mind collapsing wave function, or a relation between QM and phenomenal consciousness in general, have designed a framework in which phenomenal consciousness might collapse wave function and thus a definitive 'classical' world emerges. The framework suggested is based on IIT or integrated information theory, and also where phenomenal consciousness – qualia – is considered as 'superposition resistant'. Here we observe that Chalmers and McQueen [8] have proposed a partial quantum Zeno effect for completing their consciousness induced collapse model. Our previous framework for the emergence of the classical world naturally includes a partial Zeno effect, with trade-offs between Zeno effect and atmospheric de-coherence. We didn't have non-local mind explicitly in the previous framework.

In addition to Chalmers and McQueen, (op. cit.), Kauffman and Roli (op. cit.) have recently proposed that the human mind cannot be algorithmic, and that the capacity to find novel affordances requires a quantum mind and qualia associated with the collapse of the wave function to a single state. The next section presents evidence that humans can, in fact, collapse the wave function.

### Section 6. We Can Collapse the Wavefunction

An old idea in quantum mechanics is that mind might have something to do with "collapse of the wave function". Von Neumann proposed this, [34]. Wigner suggested the same idea at one point, see for example Wigner [35].

Following Heisenberg, as noted, we propose Res potentia, ontologically real Possibles, and Res extensa, ontologically real Actuals. Here "actualization" converts Possibles to Actuals. This assertion is fully consistent with recent results, (Gao, op. cit., Bell, op. cit.), that seem to rule out physical causes of actualization. A physical cause cannot convert a possible to an actual.

Res potentia and Res extensa plus actualization is the first new idea about mind and body since Descartes' substance dualism, Spinoza's neutral monism, Berkeley's Idealism, and pure materialism. Res potentia and Res extensa is not a substance dualism. Potentia are not substances. Thus, this view does not inherit the mind-body problem.

Instead, Res potentia and Res extensia suggest a natural role for mind. Mind "actualizes" Possibles to Actuals.

Strong evidence now supports this scientifically testable hypothesis. Radin and his colleagues, for example Kauffman and Radin [10] have tested the capacity of humans paying attention to modify the intensities of the adjacent central bands in the famous interference pattern of the two slit experiment. The effect is weak but has been tested in 30 independent experiments. At present the positive results are very strongly statistically significant at 6.49 Sigma. The probability, "p", that this arises by chance now stands a less that 4 x 100,000,000,000, Kauffman and Radin [10].

This is strong enough to take very seriously as yet further data are sought. If accepted, the results alter the foundations of Quantum Mechanics with a fundamental role for mind. Indeed, even a responsible free will is not ruled out, Kauffman Radin [10].

For the purposes of this article, we will accept these results as true.

### Section 7. Quantum Gravity if Nonlocality is Fundamental

One of us has recently published a work on quantum gravity [4]. The starting point is to take nonlocality as fundamental. Nonlocality arises in the presence of entangled coherent quantum variables. If one starts with nonlocality it is not necessary to explain nonlocality, but necessary to explain locality. Somehow locality – spacetime– is to emerge from the behaviors of the quantum variables. This immediately flatly contradicts General Relativity, which is local, and in which there is no emergence of spacetime. Further, General Relativity can be formulated in the absence of matter so matter cannot be necessary for the very existence of spacetime. But if one starts with nonlocality, the emergence of spacetime must depend on the matter – the entangled coherent quantum variables. A further note is that there is no apriori reason not to start with nonlocality as fundamental.

The steps in building this new theory of quantum gravity start with N entangled variables in Hilbert space, then constructs a metric distance between each entangled pair of variables as the sub-additive von Neumann Entropy between that pair. Sub-additive von Neumann Entropy, therefore, fits the triangle inequality. The next step notes that quantum variables can be in superposition and interpreted as potentia, neither true nor false. All the variables of classical physics are true or false. Hence the next step in the development of the theory maps distances in Hilbert space to classical spacetime distances between a succession of true actualization events. In this mapping entangled near-neighbours in Hilbert space construct themselves into nearby points in classical spacetime. The hypothesis that actualization events construct spacetime is probably testable using the Casimir effect.

### Section 8. Emergence of the Classical World

Here we refer to the ontological framework developed by Kauffman and Patra [24] which also forms one reference for the current framework, though we didn't include non-local mind in our previous work. We based our previous work on the premise that measurement and actualization, which creates the definitive classical world (this coincides with the contextuality-complementarity philosophy of Bohr[1]) can happen onlyin a specific basis. However, we still do not have a comprehensive theory for the emergence of a specific basis, except the recent attempts from Quantum Darwinism perspectives as proposed by Zurek [36] in terms of de-coherence theory. We note that decoherence does not yield a specific basis.

We have proposed the following steps for the emergence of classical world, in which testable experiments can be performed.

   i. We start with sets of N entangled quantum variables, which need not be maximally entangled. Variables can mutually actualize each other, which is approximated by the quantum-Zeno effect.

   ii. Such actualization occurs in one of the $2^N$ bases.

   iii. Mutual actualization breaks symmetry among these $2^N$ bases.

An amplitude for a specific basis can emerge and increase with further measurement in the same particular basis, it can also decay between measurements.

Here one can also refer to recent works of Kastrup [37] where if we claim that only actualization creates the definitive world, which would mean no pre-existing values, we should also accept that the world as a whole is beyond only physical, or the typical physical closure principle would not work.

   I. As the number of variables, N, in the system increases, the number of Quantum Zeno mediated measurements among the N variables increases.

   II. Now for experimental purposes, quantum ordered, quantum critical, and quantum chaotic peptides that decohere at nanosecond versus femtosecond time scales can be used as test objects.

   III. By varying the number of amino acids, N, and the use of quantum ordered, critical, or chaotic peptides, the ratio of decoherence to Quantum Zeno effectscan be tuned. This enables new means to probe the emergence of one amonga set of initially entangled bases via weak measurements after preparing the system in a mixed basis condition.

   IV. Use of the three stable isotopes of carbon, oxygen, and nitrogen and the five stable isotopes of sulfur

allows any ten atoms in the test peptide or protein to be discriminably labelled and the basis of emergence for those labelled atoms can be detected by weak measurements. We present an initial mathematical framework for this theory, and we propose experiments.

### Section 9. If Mind Is Outside of Spacetime, What Is 'My" Mind?

If we are to make sense of Aerts et. al data [15-17] and do so by proposing that mind is outside of spacetime but that the cognitive experience is in spacetime, we must claim that qualia emerge upon actualization events, as discussed above. But in addition, it becomes fundamental to address the issue: What maps the quantum variables in Hilbertspace and the vacuum to "My Mind"?

The theory of quantum gravity based on nonlocality as fundamental almost automatically affords a possible answer to this issue. Compare the relatively simple quantum behaviors of a quantum variables in a quartz crystal and the presumably far more complex behaviors of the quantum variables in the diverse proteins in a specific human brain with its unique genetic background and life experiences. The proposal is that when these quantum variables become coherent, they are not in spacetime but part of the quantum vacuum. The behaviors of these variables in the vacuum must exhibit and reflect the complexities the quantum behaviors in that specific brain. Because entangled neighbors in Hilbert space map to spatial neighbors in classical spacetime and the matter in it, actualization events with qualia will typically map to and occur in the same brain. Thus, "What is My Mind" seems naturally answered.

These proposals claim to answer Aerts et. al (op. cit.). My mind is not in spacetime, so not bound by continuity of action and nonsignaling. The Tsirelson bound can be surpassed. But actualization occurs in my brain so are my qualia.

### Section 10. The Quantum Vacuum and the Matter in the Universe

Our proposal to start with nonlocality as fundamental drives a different conception of the quantum vacuum. This vacuum is normally conceived in the absence of any matter and as a coupling of all the fundamental fields. The same can be considered as coupled quantum harmonic oscillators whose zero point energy can be studied. As so conceived, the spectrum of the quantum vacuum must be stationary.

By contrast, if nonlocality is taken as fundamental, spacetime is not fundamental and can only arise due to the behaviors of the quantum variables when coherent and also when not coherent. In the latter case, the Schrödinger equation no longer applies. The quantum behaviors of quarks, protons, neutrons, and electrons in complex proteins must differ from those in a simple crystal. With this seemingly necessary inference, the behaviors of the quantum vacuum – coherent entangled quantum variables, cannot be stationary over the history of the universe as more and more complex classical systems, stable for long periods, come into existence. We can propose, quantum vacuum must also reflect the history of the behaviours of the ever more complex matter than has come to exist and vanished.

### Section 11. The Six-Part Ontological Framework: Mind and Cosmos

The above considerations lead us to propose that: i. The quantum vacuum is composed of entangled coherent quantum variables that are ontologically real "Possibles"; ii "Mind" is identical to the Possibles of the quantum vacuum. Hence this is the definition of mind in our framework; iii. The vacuum is outside of spacetime; iv. Mind can mediate actualization of potentia, (Kauffman and Radin, op. cit.); actualization of potentia then constructs classical spacetime, where a metric exists in the quantum vacuum Hilbert space via non-additive von Neumann Entropies between pairs of entangled variables, that is then mapped to events at specific classical spacetime locations, (Kauffman quantum gravity); v. We experience such actualized quantum variables as "qualia" [8, 9]; vi. In the last, sixth, part we propose the emergence of classical world, which is based on our previously proposed framework [24]. We suggest a mutual actualization process of quantum variables. Through a trade-off between the quantum Zeno effect and atmospheric de-coherence, such de-cohering and re-cohering variables creates the observable classical variables. In this framework we suggest verifiable experiments with peptides whose entangled variables decohere exponentially fast versus peptides whose entangled variables decohere power law slowly as a possible ground of test.

We hope to show that the six part proposal above allows us to account for Aerts' et. al results that surpass the Tsirelson bound. Far more, this new six part framework may help organize our emerging ideas about "Mind, Matter, and Cosmos".

### Section 12. Discussion and Further Work

Nonlocality has always baffled us. The non-local and non-deterministic collapse of wavefunction in QM worried Einstein throughout his working life, since the fear was such nonlocality would mean action at a distance and thus break- down of the causality structure of spacetime. The latter is fundamental to any Physical theory. Certainly, a huge literature has demonstrated that non-local collapse may not mean any superluminal signaling. Later since Bell's seminal contribution, there have been many versions of such frameworks (CHSH being the most popular), which have suggested that local hidden variable theories cannot reproduce QM faithfully. Loophole free Bell inequality/ CHSH inequality violations have demonstrated that one or more of the basic underlying assumptions, of localism, realism or non-contextuality, or statistical independence have to be relaxed to describe the empirical results of QM.

Many workers have shown that Bell inequalities violations are considered to be evidence of non-local correlations between subsystems. The canonical example is entangled pairs of particles (EPR set up for example) with agents measuring on each half of the pair who are space like separated.

In the presence of an assumed background spacetime, the only way a no-signaling theorem is going to be preserved is by introducing inherent quantum uncertainty in outcomes. In the words of Shimony there can be a happy co-existence between Special Relativity and quantum fundamental uncertainty. However, the Hilbert space structure, assumed to be the state space in such frameworks, inherently does set up an upper bound for violations of inequalities, the celebrated Tsirelson (or Cirelson) bound. Thus, the question arises: What, if any, empirical observations exist where such a limit is violated?

Over the last decade there has been strong evidence of violations of CHSH inequalities, pertaining to cognitive experiments (Aerts et. al). The data are now confirmed at p = .001 to .005. Further work is needed to confirm these results more strongly. However, they are already strong enough to warrant consideration of the implications.

Aerts et. al (op. cit.) have tried to preserve a background spacetime and "no-signaling" by assuming more complex entanglement, i.e. both states and measurements. The same authors have also claimed that quantum entities might be conceptual or cognitive entities, hence non-spatial.

We propose here a novel, yet unexplored framework based on non-localism, where spacetime need not be fundamental to existence. Nonlocality is not mysterious in our framework. Our attempt is to start from nonlocality and derive locality from first principles. Then in such a constructed local spacetime we have standard QM and SR operate with restricted nonlocality, which is no-signaling also.

Our proposal is related to that of Aerts et. al in an unexpected way. As just noted, these authors propose quantum entities might be conceptual or cognitive entities, hence non-spatial. Almost in parallel, we propose that the quantum vacuum consists in ontologically real Possibles, that Possibles are non-spatial, i.e. not in spacetime, that Mind is identical to these Possibles, that Mind can actualize these potentia, and we can experience these as qualia.

What should we make of this extensive new six-part framework? A first point is that other attempts such as PR boxes correspond to no known physical reality.

Our proposal is not too distant from Aharanov's nonlocality proposal. But as Shimony notes, this is "passion at a distance" in spacetime. In our six part framework, the correlations are among ontologically real possibles that are not in spacetime, but "mind" is/are part of the quantum vacuum of possibles. These possibles then constitute the information Zeilinger hopes is the basis, somehow, of spacetime. However, Zeilinger offers no account of what information is, other than a "bit", nor any idea of how these might be related to spacetime.

In our account, spacetime is constructed by the sequential actualization of quantum variables in Hilbert space with a metric via non-additive von Neumann Entropies that then map to Actual events whose mutual distance relations reflect the metric in Hilbert space to constitute spacetime [4]. This claim underlies our first part, "i", and "iii". There are data at 6.49 sigma to support "iv" and "v" above, (Kauffman, op. cit.).

The second part, ii "mind" is identical to the possibles of the quantum vacuum, is an entirely new proposal. Oddly, this proposal just might afford a highly speculative answer to the point raised in a recent article on Biocosmology [38] about a link between the emergence of life 4 billion years ago and the recent dominance of dark energy whose tight temporal coincidence in Cosmology is strange. If living organisms actualize quantum variables far more often than the quantum variables of the abiotic universe, then life, via mind, can have played a role in the emerging dominance of dark energy in the past four billion years.

Our vi. part concerns the emergence of the classical world from the quantum world. Our own proposal [24] has the virtue of being testable. In addition, it automatically supplies the incomplete Quantum Zeno Effect desired by Chalmers and McQueen, (op. cit.). Our specific proposal for the emergence of the classical world is consistent with our general framework i. to vi. and is consistent with efforts to study how an increase in the mass of molecules such as the Buckyball may increase decoherence.

The proposal that quantum gravity is a quantum construction of spacetime is not yet united with General Relativity but may be a new pathway to do so. Such a union with our proposals in the present article might be fundamentally new. Such a union would embrace Mind, Matter and Cosmos.

The most important lines of further work are: i. Experiments to test and extend the Aerts et. al results, (op. cit.). A 'p' value of 0.001 is of interest, but hardly persuasive. ii. Our six part framework rests heavily on taking nonlocality as fundamental. Experiments testing the hypothesis that actualization constructs spacetime are needed. The Casimir effect may prove useful [4]. iii. Further testing of the capacity of the human mind to 'collapse the wave function' are needed. The current data at a Sigma of 6.49 are strong. But this is a truly major claim that must pass muster with critics. iv. Were it possible to demonstrate that actualization events constructed spacetime and with good grounds established that mind could mediate actualization, it might become possible someday to test if mind by mediating actualization can construct spacetime. v. The current article is at best a conceptual framework. A far more formal and integrated mathematical theory must be constructed and ultimately tested.

## Conclusions

The dream of physics since the discoveries of General Relativity and Quantum Mechanics nearly a century ago has been their union in Quantum Gravity. Yet since Newton, a role for mind in the becoming of the Cosmos has seemed precluded. In 2022 NASA launched a rocket that nudged a distant asteroid in the Solar System into a slightly different orbit altering the orbital dynamics of the solar system. Mind has cosmic consequences.

In the current article we take the results of Aerts et. al as if they were firmly established. With a p value of .001

the results are at most grounds for consideration. More experiments are needed. However, assuming such firm results, human cognitive events surpass the Tsirelson bound. Attempts to explain such a result within a background spacetime that demands Continuity of Action and no-signaling have severe difficulties.

Were mind not in spacetime the requirement for Continuity of Action and nonsignaling would not arise. The possibility of cognitive events surpassing the Tsirelson bound would arise. But this would require that mind correspond to something "real" that is not in spacetime, and also that cognitive events themselves exist in the actual experienceof humans, hence in spacetime.

We approach quantum gravity by taking nonlocality as fundamental. If nonlocality is fundamental, spacetime is not fundamental. Non locality arises with two or more entangled coherent variables. Thus, we are forced to the conclusion that spacetime somehow emerges from the behaviors of coherent entangled variables. This flatly contradicts General Relativity, which is local, spacetime does not emerge in General relativity, and GR can be formulated without matter fields so the very existence of spacetime cannot depend upon matter.

Based on the above it is straightforward to define a metric distance between each pair ofentangled variables in Hilbert space as the subadditive von Neumann Entropy of that entangled pair. But this set of distances is in Hilbert space whose variables can be in superposition. Classical events in spacetime cannot be in superposition. Hence it becomes natural to find a map between the metric in Hilbert space and classical spacetime by successive actualization events in which the distance between actual events correlates with those of the corresponding variables in Hilbert space. Nearby entangled variables in Hilbert space construct themselves into nearby points in classical spacetime.

In the present article we hope to account for evidence that cognitive events do surpass the Tsirelson bound by identifying mind with coherent entangled quantum variables that constitute the quantum vacuum and are not in spacetime. These are to map to cognitive events within spacetime by actualization events that constitute qualia.

Increasingly, strong grounds exist to support the view that conscious events – qualia – accord with actualization events. Further, evidence now stands at 6.49 sigma, or 4 in 100,000,000,000, in support of the claim that mind acausally mediates actualization.

The union of the above issues then constitute a vision of quantum gravity in which mind is identical to the entangled coherent quantum variables of the quantum vacuum and mind itself mediates the actualization events that construct classical spacetime.

Such a vision is not yet united with General Relativity. A new union may be possible in which quantum gravity constructs the classical spacetime in which General Relativity operates.

General Relativity requires a world of classical objects. Among these, some are very simple, some like the evolved proteins in the human brain are very complex. The quantum behaviors of very complex molecules and groups of molecules will be far richer than those of a simple small quartz crystal. Therefore, the mind of a brain can be far more complex that a mind of a crystal. And the quantum behaviors of one brain will be partially unique to that brain and its ontogenetic and experiential history. But the quantum behaviors of entangled variables in brains, when coherent, are not in spacetime and are part of the quantum vacuum. Upon actualization, these entangled variables that are neighbors in Hilbert space construct themselves to nearby points in the matter in classical spacetime, thus typically to events located in the same brain. "My memories and thoughts are mine, not yours." Yet by entanglement between brains, telepathy is possible, and precognition is possible. By entanglement between variables in a brain and other physical objects, psychokinesis is possible. The data for all these are now abundant at high Sigma values.

We wish to make a converse point. If the data we analyze require that we abandon spacetime and its matter and energy as "all that exists", this is the first experimental evidence that something real exists beyond spacetime.

The concepts and data we have discussed do not yet warrant such enormous conclusions. Far more would be required. Yet, perhaps for the first time since Newton, they may constitute the start of a conceptual framework uniting Mind, Matter and Cosmos.

**Appendix**

Entanglement, nonlocality and usage of Bell-type inequalities outside physics

We would like to refer to the emerging literature of usage and implications thereof for Bell type inequalities outside Physics. In recent cognitive science literature [(Aerts et al (2022) or Dzhafarov et al (2022)] there has been wide usage of Bell type inequalities, with claims of Contextuality and Nonlocality in cognition. Aerts et al have produced the most prolific of these works, as referred to above. In the works of Aerts et al, or 'Brussel's' group approach to human cognition, experiments have been suggested based on how concepts are combined in human mind, for example animal names and the sounds they make. It has been proposed that such 'non-separable' states violates classical correlation limit, as would be suggested by the non-violations of Bell type inequalities, in some cases a super quantum violation or crossing the 'Tsirelson' limit have also been observed. However here we remind ourselves that such experimental results need to be met with caution, since the nature of 'entanglement' is not clear in domains outside physics; hence, here we observe some points, which might be insightful.

In quantum physics literature since the EPR work [18] entanglement among 'space-like' separated subsystems of a composite system has been studied numerously[one should also incorporate the Nobel Prize in Physics in 2022 to the experimental pioneers in testing Bell inequality violations for such space-like separated subsystems]. Space-like separation is critical for particle physics case, since otherwise it would be trivial to conclude that there are direct influences between such sub-systems, which might not have anything to do with 'entanglement' or subsequent Bell inequality violation. In other words physicists would need a 'no-signaling' experimental set up, such that when measurement is performed on one such separated sub-system, before observing the result one should not get any signal from the observation on the other subsystem measurement. Such no-signaling is according to relativistic causality, which demands that there is an ultimate finite speed of signaling between events in Spacetime, which is the speed of light in vacuum.

Given this set up Bell type inequalities are constructed [ we have explicit formulations shown in papers on modelling decision making], and the main proposition is that under no-signaling, such inequalities should have an upper bound of absolute value 2, certainly algebraically the violations can be up to 4, the physical constraint ties down to the upper limit of 2. However, in real quantum physics experiments with entangled pairs of particles, such an upper limit of 2 is violated, and a new upper limit of $2\sqrt{2}$ is suggested, Tsirelson limit by name. This result has been verified numerous time, which is the main source of speculation about Nonlocality in quantum physics. Certainly, the experiments have to be loop hole free, which mainly means no-signaling condition has to be maintained.

However, here we further emphasise that Bell type inequalities are rather general in nature. Hence, it is legitimate to use it in other domains of knowledge too. More specifically, Sandu Popescu [12] and his collaborators have reframed EPR type experimental setting into generalised game setting, such alternative framework has been called as PR boxes. Again, in classical electromagnetism, where no quantum degrees of freedom is conceived, and Maxwell's field equations are used, there also we have intra system entanglement states which violates Bell type inequalities. More recently, entanglement has been viewed from Neumann entropy measurement. Neumann entropy, is a measure of 'purity' of states, which would mean whether the 'composite' entity is more close to pure state as defined in quantum mechanics (ray in the respective Hilbert space) or a mixed state (which is an epistemic probability distribution over such pure states). The formulation shows that for maximally entangled states the entropy measure is 0, whereas the subsystems are always in mixed states, hence with larger entropy. This is the sub-additivity of Neumann entropy measure. It can be proposed that entanglement can be viewed as a preparation process where sub systems collaborate in such a way as to reduce entropy in the composite state, hence this definition might not require the standard measurement picture on space-like separated sub-systems. Hence a few points can be observed about nature of entanglement and nonlocality, Contextuality as below.

1. Concept of entanglement need not be grounded in standard space-like separated subsystems only, as has been the central focus of quantum physics or particle physics. If entanglement is thought as to be a process where composite systems entropy is lower than subsystems, where the composite state is ideally a pure state, but sub-systems are in a mixed state, then we might observe such structures in other domains of knowledge too, for example in consciousness or more simply in cognitive experiments. Classical entanglement concept in Maxwell's field equations based optics is well known, and referred to in the current paper also.
2. Nonlocality can be a more general concept, even without entanglement, based on indistinguishability of quantum particles. Some authors have suggested mapping between nonlocality and Contextuality via violations of Bell inequalities.
3. We propose rather a general framework based on nonlocality principle, not necessarily grounded in the relativistic physical Spacetime background. Hence mind/cognition might be appropriate domains to work with.